# Resonant dynamics of one-side multipactor on dielectric surface

Gennady Romanov[1]

*Fermi National Accelerator Laboratory, Batavia, IL, USA*

(*Electronic mail: gromanov@fnal.gov)

(Dated: 26 May 2023)

Breakdown of dielectric RF windows is an important issue for particle accelerators and high-power RF sources. One of the common reasons for RF window failure is the multipactor on a dielectric surface. The multipactor may be responsible for excessive heating of the dielectric and discharge of charges that accumulated in the ceramic due to secondary emission. In this study, comprehensive self-consistent PIC simulations with space charge effect were performed. This was to better understand the dynamic of one-side multipactor development and floating potential on the dielectric induced by the emission. The important correlations between the multipactor parameters at saturation, the secondary emission properties of dielectric, and the applied RF field parameters have been found, which led to the conclusion that the dynamics of one-side multipactor on dielectric is a resonant phenomenon.

## I. INTRODUCTION

One side multipactor with the RF field parallel to the dielectric surface, which is typical for RF windows, requires a returning force to develop. In the case of isolated metal or dielectric body the returning force can be the result of a floating potential which is due to the charging of the isolated body by emission current. Also, an inhomogeneous RF field can by itself ensure the return of the emitted electrons to the body surface[1], but only a returning force due to dielectric charge will be considered here. Buildup of surface charge starts with random collisions of electrons that come from other processes and sources with sufficient energy to generate a number of secondary electrons. If certain conditions are met, then, at early stages of multipactor development, the emission current (the secondary electrons that leave the body) is larger than the collision current (the electrons that return to and hit the body), so the surface charge buildup continues, and positive electric charge is accumulated on the body. With increasing returning force more secondary electrons with higher initial energy start to return to the emitting surface and contribute to the floating potential. This stochastic process requires a material with sufficiently high secondary emission yield (SEY) to be realized, and, unfortunately, the dielectric materials used for RF windows typically have very high secondary emission yield (SEY=8-10 for alumina). Obviously, this charging process cannot continue indefinitely, and eventually the process comes to saturation with some equilibrium floating potential on the dielectric[2]. Most studies on one-side multipactor discharge where the RF field is parallel to the surface were performed analytically and or by numerical simulations based on simplifying assumptions[2-4]. Nowadays the simulations are performed more often with the particle-in-cell method (PIC) which is gradually becoming a primary tool for multipactor study[5-7]. An ultimate advantage of the PIC method is that the method employs the fundamental equations without approximation, allowing it to retain most of the physics[8]. In this work the one-side multipactor on dielectric was studied in detail with self-consistent particle in cell (PIC) numerical simulations using CST Particle Studio[9]. The main features of this PIC code are true 3D multiparticle dynamics, space charge effects, self-consistent RF and static fields, advanced secondary emission models, and well developed post-processing. Collective effects also play a significant role in multipactor dynamics in general, and here they are especially important. The capability to track millions of macroparticles randomly distributed over energies and phases was a decisive factor in the case under study. Earlier simulations of saturated multipacting on dielectric surface performed with the CST PIC solver clearly demonstrated a synchronous motion of the particle ensemble with the RF field and indicated other signs of a resonant character of the process[10]. This manifest contradicted the common thinking, in which case this multipactor is seen as a stochastic phenomenon. That fact has required further study to evaluate the hypothesis of a resonance dynamics in this type of multipactor.

## II. PARTICLE-IN-CELL MODEL

The principal PIC model is simple: it is a dielectric plate 40x20x0.2 mm placed in static and radio-frequency (RF) electric fields (Fig.1). The uniform electrostatic field is perpendicular to the plate surface and acts as a returning force in the simulations without space charge effect, but it is disabled in the simulations with space charge effects. With space charge effect a returning force is generated by a positive charge accumulated on the dielectric plate. The uniform RF electric field is parallel to the dielectric surface in both cases and provides the electrons with energy for secondary electron generation. The equations of electron motion in this case are as follows:

$$m\ddot{y} = -eE_{dc}; \quad m\ddot{x} = -eE_{rf}\sin(2\pi ft + \theta) \quad (1)$$

where $x$ and $y$ are respectively horizontal and vertical coordinates of the electron; $m$ – electron mass; $e$ – electron charge; $E_{dc}$ – static electric field (external or induced by MP); $E_{rf}$ – amplitude of RF electric field; $f$ – frequency of the RF field; $\theta$ – phase of the RF field at the moment of electron emission (initial phase of an emitted particle).

The emission property of the plate material is defined by an assigned secondary emission model. The CST PIC library employs the advanced probabilistic Furman-Pivi emission model, which includes elastic and diffusion emissions, and was not used in this PIC model. This was due to the fact that the simulations were performed mostly with GPU acceleration, which currently works only with imported true emission models. Additionally, there are no reliable data on elastic and diffusion emissions for RF ceramics. Because of these two reasons the imported Vaughan emission model was assigned to the dielectric plate. The general SEY function from the model is shown in Fig.2. The maximum of the SEY function was varied in the simulations from 1.5 to 3, which is much lower than in reality. The SEY was lowered in the simulations to avoid excessive number of particles to track and reduce simulation time. The initial energy of secondary electrons $W_0$ in this and all other emission models is determined by a gamma distribution. The probability density function of the distribution is also shown in Fig.2. The azimuthal scattering angle is uniformly distributed, whereas the probability distribution function for the polar angle is given by $f_\alpha = \cos\alpha$, $0 < \alpha < \pi/2$. The secondary emission parameters were variable during the simulations and their exact values are given in the text where it is relevant.

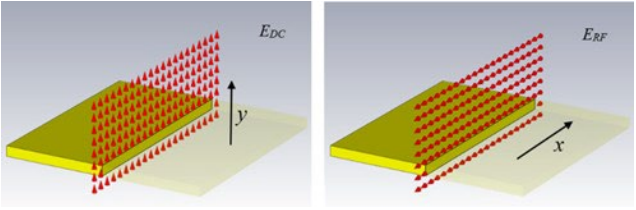

FIG. 1. Cross-sections of electrostatic and RF field distributions.

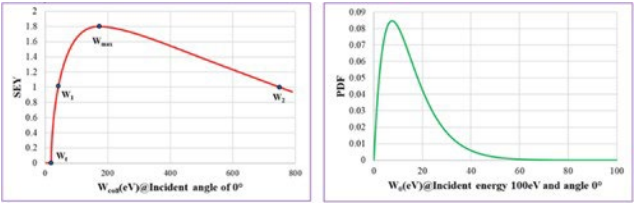

FIG. 2. The general Vaughan SEY function and the probability density function of initial energy of secondary electrons. In the SEY plot the important incident energies are marked: threshold energy $W_t$, first crossover $W_1$, SEY maximal $W_{max}$ and second crossover $W_2$. In PDF plot the important point is the most probable energy of emission $W_{0max}$ (or temperature of emitted electrons $T_e$)

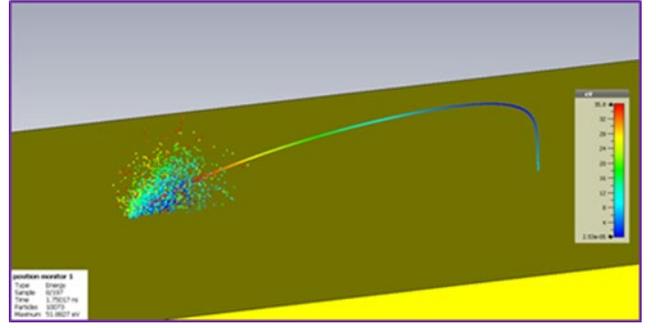

FIG. 3. Particle distribution at $t=T/2$ after start of emission at zero initial phase. The curve before collision is not a particle trajectory, but a continuous chain of particles. After the collision there is a cloud of the secondary electrons with random initial energies and directions according to SEY model.

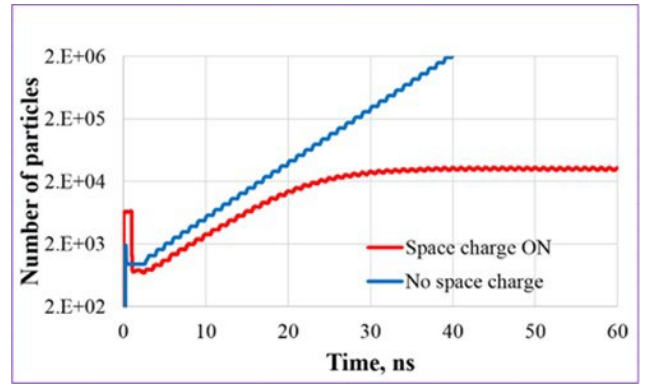

FIG. 4. Saturation of number of particles in the PIC simulations of multipactor with space charge ON compared to the simulations in the same model without space charge effect (vertical scale is logarithmic).

The source of initial particles was placed in the center of the plate. It emitted particles at start of the simulations during one to two RF periods ($T=1/f$) to cover all possible initial phases of particles. For easier interpretation of the beginning stages of multipacting, the initial electrons were mono-energetic with a fixed energy of 7.5 eV and did not have an angular spread – they all were emitted perpendicularly to the surface. Note, that this setting applies to initial particles only – after injection in the following simulations the parameters of secondary electrons were governed by the chosen emission model described above (see Fig.3).

The growth of particle number in the multipacting process is different in PIC simulations performed with and without space charge effect: the number of particles exponentially increases in simulations without space charge effect, but it saturates when space charge effect is considered (see typical plots in Fig.4). Accordingly, the parameters used to indicate multiplication are different with and without space charge effect.

The main parameters used for indication of multipacting (MP) and evaluation of its intensity are emission/collision currents (the macroparticles have electric charge, so the currents can be calculated even when space charge effect is not considered), effective secondary emission yield <SEY>, and energy of collision, all defined as:

$$<SEY> = \frac{I_{\text{emission}}}{I_{\text{collision}}}; \quad W_{\text{collision}}(eV) = \frac{P_{\text{collision}}(W)}{I_{\text{collision}}(A)} \quad (2)$$

where currents $I_{\text{collision}}$ and $I_{\text{emission}}$ and collision power $P_{\text{collision}}$ are standard CST PIC output parameters averaged over the last 3-5 RF periods (typical simulation times are 30-100 RF periods). In the simulations without space charge effect, <SEY> exceeding 1 indicates multipacting, and the higher its value the more intense multipacting is, which is also indicated by $I_{\text{emission}}$. With active space charge effect the MP process saturates making <SEY> = 1 since $I_{coll}=I_{emis}$ at saturation. So, the absolute value of collision (or emission) current is the only indicator of multipacting intensity in the simulations with space charge effects. The parameters (2) characterize the whole process, but it is also possible to analyse behaviour and dynamics parameters of individual particles using post-processing templates.

### III. SIMULATION WITHOUT SPACE CHARGE

The simulations without space charge effect do not show the correct particle dynamics but they help to define the range of parameters and reveal important correlations between them, which helps to understand resonant particle dynamics concepts and floating potential development on the dielectric. The simulations were performed for a broad range of parameter values, but qualitatively the results were very similar. Therefore, for simplicity a typical set of parameters is considered initially, and some generalization will be presented later on.

The parameters of the dielectric emission model were: maximal $SEY_{max}=1.5$ at $W_{max}=150$ eV, $W_t=0$, $W_1=33$ eV and $W_2=667$ eV. The most probable energy of 7.5 eV was set in the probability density function of initial energy of secondary electrons (CST default). It is obvious that the initial energy of secondary electrons cannot be zero, because otherwise no emission could exist (though this assumption works fine in cases where the RF field is perpendicular to the emitting surface). But it is vital to use a realistic continuous gamma distribution for the secondary electron energy, because that provides a foundation for the resonant character of this type of multipactor. Indeed, in perfect field distributions (which means there is no RF field component perpendicular to the dielectric surface) a time-of-flight t is defined only by electrostatic field and initial velocity of secondary particle:

$$t = 2\frac{m\ v_0}{e\ E_{dc}} = T/2 \quad (3)$$

where $v_0$ is the initial velocity of secondary electron. From this simple relation, it follows that for any electrostatic field level there are secondary particles from a continuous infinite energy spectrum with an initial velocity such that that they would fly exactly half an RF period between emission and collision. In other words, there are always resonant particles among secondaries at any level of electrostatic field.

The frequency of the external RF field was 325 MHz, one of the main subharmonics used in modern superconducting linear accelerators. The field levels at the start of study were analytically estimated using equations (1) in the following way. The initial amplitude of the RF field was set 17.4 kV/m, which provides a resonant secondary electron with energy up to $W_1=33$ eV during half an RF period and at phase of emission $\theta = 0°$. The electrostatic field level was chosen $E_{dc} = 12$ kV/m to make the time of flight of the secondary electron with the most probable emission energy of 7.5 eV equal to half of RF period T/2. Starting from these initial field levels the simulations with variable $E_{dc}$ and $E_{rf}$ were performed to evaluate the effect of the electric fields and define their threshold levels. Fig.5 shows the number of particles vs time obtained in simulations with fixed electrostatic field level $E_{dc} = 12$ kV/m and variable RF field amplitude $E_{rf}$. Fig.6 shows the average collision energy and effective secondary emission yield obtained in the same simulations. The multipactor here starts at $E_{rf} \approx 26$ kV/m, which is much higher than the initial $E_{rf}=17.4$ kV/m. The reason is that at this initial RF field level $E_{rf}$ only the resonant particles with an initial energy 7.5 eV and emitted at initial phase = 0° can gain energy equal to the first crossover $W_1$. Most of the secondaries generated by the resonant particles are not resonant due to the initial energy spread. Low energy non-resonant particles cannot gain $W_1$ because their time of flight is shorter than T/2. High energy non-resonant particles can gain sufficient energy for re-emission, but they fly longer and collide less frequently. Therefore the average collision energy of the particle bunch at this initial RF field level is well below $W_1$. As a result, the multipacting process starts at much higher RF electric field than the simple estimation based on single resonant particle dynamic predicts.

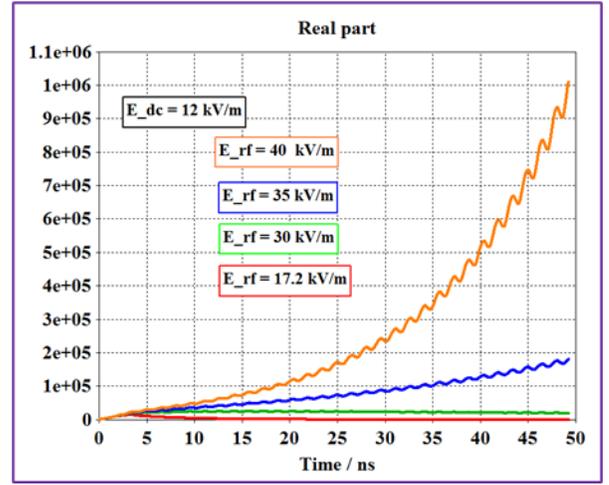

FIG. 5. Number of particles vs time at different RF field amplitudes and constant electrostatic returning field of 12 kV/m.

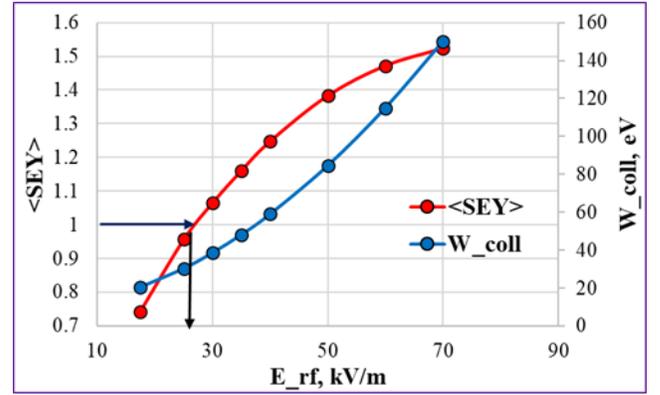

FIG. 6. Average collision energy and effective secondary emission yield as functions of RF field amplitude at electrostatic field of 12 kV/m. The arrows show the breakdown level of RF field.

On the other hand the presence of even a small number of resonance particles does matter. Evaluation of the breakdown level of the RF field for very low electrostatic field made in[11,12] assumes only non-resonant motion of the electrons (so called polyphase regime: the time of flight T for all electrons, the collision phases are uniformly distributed over the RF period). Following these works the estimation of the breakdown level of $E_{rf}$ given the emission parameter $W_1$ = 33.5 eV and f=325 MHz is:

$$E_{rf\ breakdown} = 2\pi f \frac{m}{e} \sqrt{2W_1/m} \approx 38 \quad \text{kV/m} \quad (4)$$

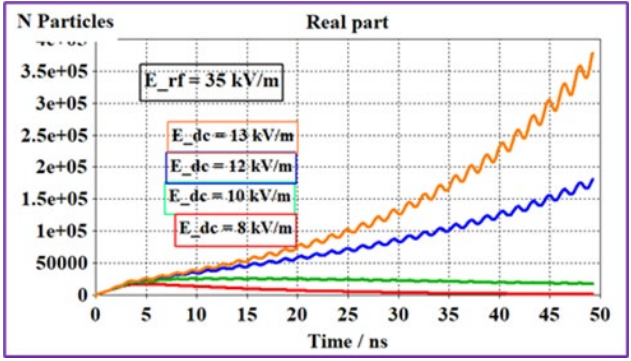

FIG. 7. Number of particles vs time at different levels of constant electrostatic field and RF field amplitude of 35 kV/m.

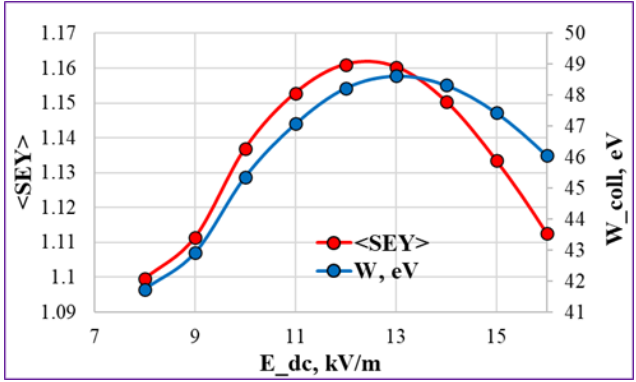

FIG. 8. The energy of collision $W_{collision}$ and effective secondary emission yield <SEY> as functions of electrostatic field at RF field amplitude of 35 kV/m.

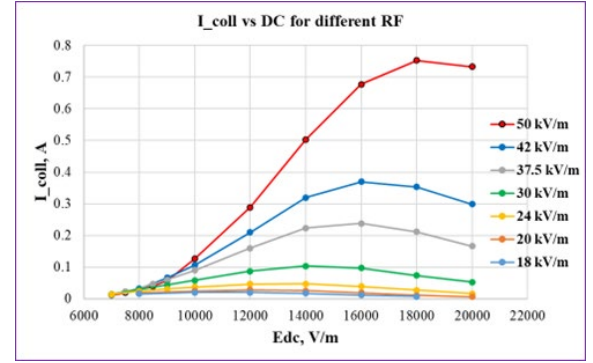

FIG. 9. Collision current as a function of the electrostatic field $E_{dc}$ at different levels of RF field amplitude $E_{rf}$.

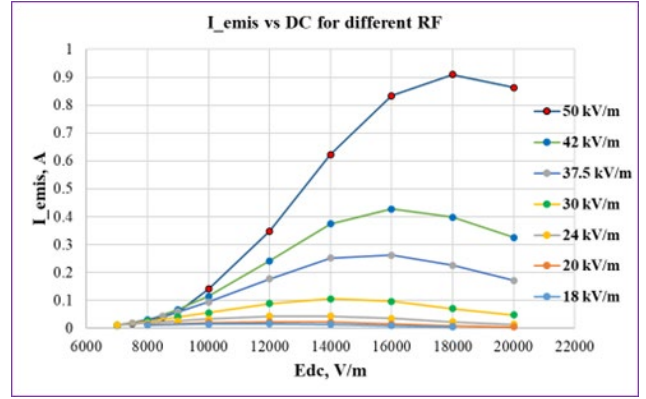

FIG. 10. Emission current as a function of the electrostatic field $E_{dc}$ at different levels of RF field amplitude $E_{rf}$.

The breakdown level of $E_{rf}$=26 kV/m obtained in the simulations is much lower, which suggests a contribution from more effective and fast resonant multiplication.

The results of simulations with a fixed RF field $E_{rf}$ = 35 kV/m and variable $E_{dc}$ are shown in Fig.7. The amplitude of $E_{rf}$ was set higher than its threshold to assure powerful multipacting. The plots demonstrate good general agreement with the simple estimation based on (3), though the actual electrostatic field threshold is lower due to the increased $E_{rf}$ and approximately equal to ≈ 8 kV/m, as can be found from Fig.8. In Fig.8 the energy of collision $W_{collision}$ and effective secondary emission yield <SEY> show non-linear dependence on $E_{dc}$ with a maximum at the level close to the "resonant" value of 12 kV/m. This non-linear dependence on $E_{dc}$ with extremes is a common feature of this type of MP and plays an important role in its dynamics.

More statistical data were collected from the simulations, where the RF field amplitude was swept from 18 kV/m to 50 kV/m, and the electrostatic field was changed from 7 kV/m to 20 kV/m. The data are presented in Fig.9-10. Simulation time was 15 RF periods; this relatively short time was chosen to reduce overall simulation time. It created an infeasible situation in which a collision current $I_{coll}$ is not equal 0 at <SEY> less than unit. The particles from the initial bunch do not have enough time to get lost and disappear as expected because of unfavourable conditions, instead they continue colliding with the plate. So, it should be kept in mind that below <SEY>=1 the collision current $I_{coll}$ and collision energy $W_{coll}$ go to zero after enough time elapsed. Despite this flaw the averaging of the parameters performed over last 5 RF periods produces smooth informative curves.

The dependence of the average parameters on $E_{dc}$ plotted in Fig.9-12 show that MP intensity declines with increasing of electrostatic field. It was mentioned above that there are always resonant secondary particles emitted with a proper initial velocity at any level of electrostatic field, and the higher electrostatic field then the higher this velocity and corresponding emission energy. But the probability of high energy particle emission, and therefore their number, drops above $T_e$ accord-ing to the given PDF shown in Fig 2. The drop of the number of resonant particles results in the reduction of the total emission current (Fig.10). Following these speculations it was logical to assume a dependence of the MP parameters on the initial energy distribution of the emitted particles, that is the probability density function of initial energies. The next set of simulations without space charge effects was performed with different temperatures of emitted electrons $T_e$ (see Fig.12); the RF field strength was 37.5 kV/m for all runs.

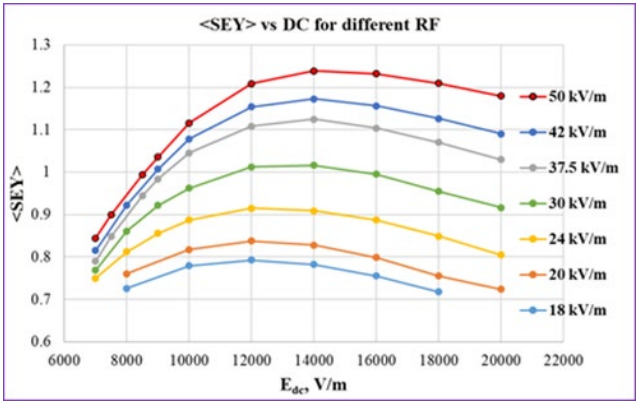

FIG. 11. Effective secondary emission yield as a function of the electrostatic field at different levels of RF field amplitude $E_{rf}$

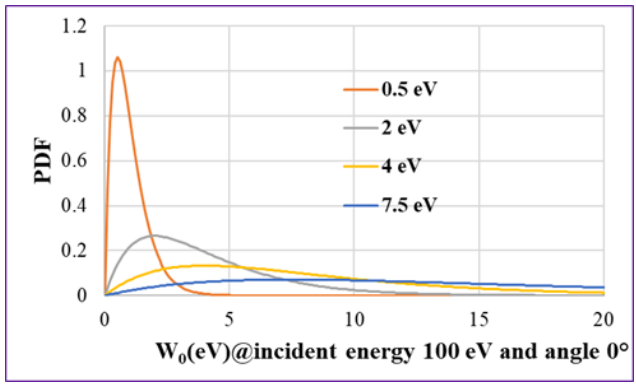

FIG. 12. PDFs of initial energy of secondary particles with different most probable values $T_e$

The collision currents and effective secondary emission yields, shown in Fig.13, have maximums correlated with the most probable initial energy $T_e$. The absolute values of <SEY> maximums are almost on the same level, which reflects the same RF field strength. The collision current is higher for higher $T_e$, which is due to the increasing intervals of initial acceptable energies[12] since PDFs are smoother for higher $W_{0max}$. The collision energy shown in Fig.14 has clear maximums correlated with the most probable initial energy $T_e$ as well. Particle velocity is a vector sum of velocity components parallel and normal to the dielectric surface. Therefore, the collision energy increases with $T_e$ simply because the vertical component of particle velocity at the moment of collision also increases with $T_e$ since it is equal to initial particle veloc-ity by magnitude (with reversed direction).

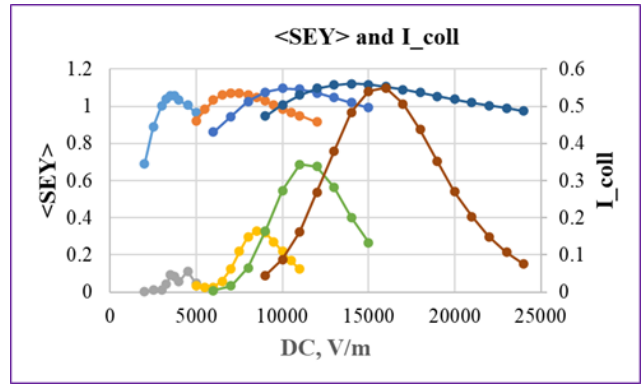

FIG. 13. Effective secondary emission yield and collision current as functions of electrostatic field for different PDF from Fig.12.

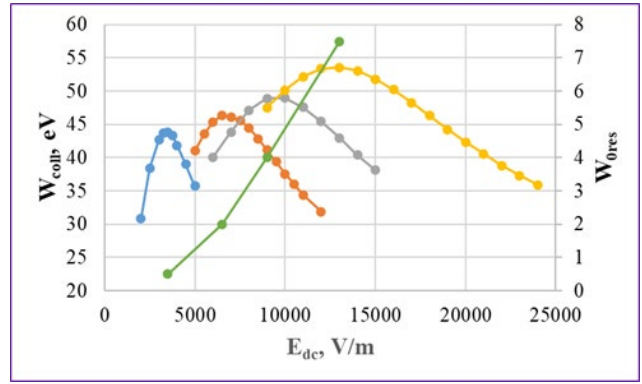

FIG. 14. Collision energy $W_{coll}$ as a function of electrostatic field for different PDF from Fig.12. Resonance initial energy $W_{0\,res}$ is shown in green.

## IV. SIMULATION WITH SPACE CHARGE EFFECT

### A. Model

For simulations with space charge effect the model was modified. The external electrostatic field was removed, and a voltage monitor (2 mm long) was added as shown in Fig.15. The single point particle source was replaced with a circular one to make the initial charging of the ceramic more uniform. Total charge emitted during one RF period was chosen to be 1e-9 C. The Vaughan emission model parameters at the start of study were $W_i=0$, $W_1=11$ eV, maximal SEY of 3.0 at $W_{max}=200$ eV and $W_2=6470$ eV.

The initial emission from the particle source instantly generates a charge on the dielectric surface, so there was no need to apply any ancillary electrostatic field, which was used in some models to initiate multipactor process [3]. The spatial particle distribution after 1.25 ns of emission is shown in Fig.16. Some particles leave the dielectric along almost straight trajectories. They are the very first particles emitted when the electrostatic field is not strong enough yet to return them back to the surface. Gradually the emitting electrons build up a positive charge on the dielectric and the particles start to return to the dielectric surface producing secondary emission.

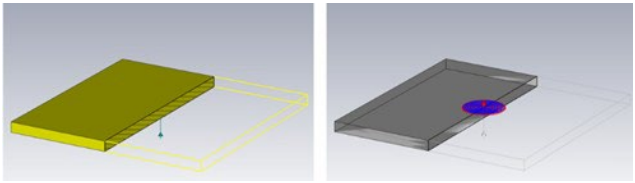

FIG. 15. Left – location of the voltage monitor is indicated by arrow. Right –the circular source of initial particles

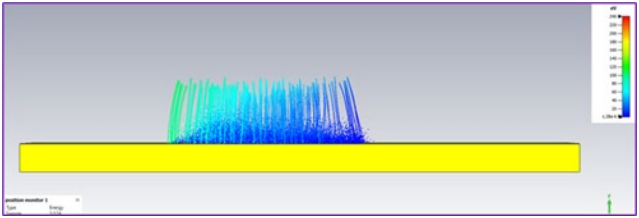

FIG. 16. Particle distribution at 2 ns after start of emission with space charge effect.

### B.  Discharge saturation. Floating potential.

The main impact of the space charge on virtually all kinds of multipactor is the saturation of the multipacting process. In the present case where the RF electric field is parallel to the dielectric surface, the saturation of the discharge follows the saturation of the charge accumulated on dielectric[2].

Development of MP at different levels of RF field during a simulation time of 15 RF periods is shown in Fig.17. Due to the lower $W_1$ = 11 eV the breakdown level of $E_{rf}$ is about 25 kV/m, which is slightly lower than what was found in the simulations without space charge effect with external electrostatic field. Further increasing of the RF field above the breakdown level increases the speed of MP development and the number of particles at saturation.

A principal difference in simulations with space charge effect is that the electrostatic returning field is generated by MP

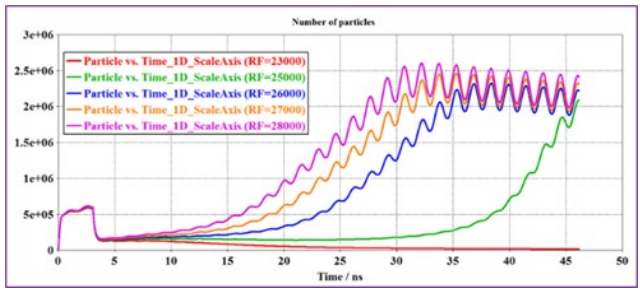

FIG. 17. Number of particles vs time at different levels of RF field.

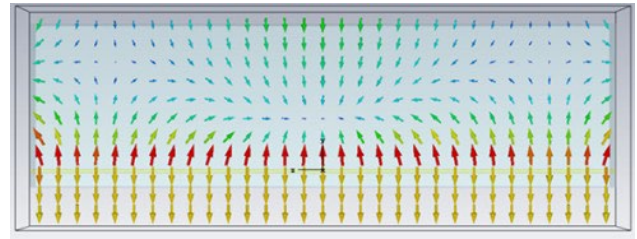

FIG. 18. Snapshot of electric field distribution.

itself and its level is regulated by MP intensity. It is easy to speculate from Fig.13-14 on the auto-regulation of the positive charge on the ceramic in the case of active space charge effect in which the charge is induced by emission. After the initial emission happens, if RF field is high enough to provide emission current higher than collision one, the charge on the ceramic increases. In the stronger returning field more energetic and more numerous secondary electrons become resonant making emission more intense and resulting in further charge accumulation. The emission current and consequently the charge and its electrostatic field increase until the secondary particles with initial energy greater than $T_e$ become resonant. But the number of secondary particles emitted with energy higher $T_e$ drops beyond the maximum of PDF and as does the total emission current. The electrostatic field decreases following the drop of emission current, and the resonance returns to the particles with lower initial energy below $T_e$.

The electrostatic field voltage monitor is located under the plate and integrates the electrostatic field along a 2 mm line perpendicular to the plate. The location has been chosen to avoid interference of the monitor with the space charge of the particle cloud. The field strength is obviously different above and below the plate (see Fig.18), and the field distribution is highly non-uniform because of the space charge of the particle cloud, so the monitor readings are relative and just indicate the general level of the returning field. The voltage monitor record in Fig.19 also shows some dependence of the saturated electrostatic field level along with dependence of growth rate on the applied RF field. When the RF field is below the breakdown level the multipactor does not develop and the plate is not charging/discharging, though the constant electrostatic field shown in Fig.19 for a low RF field of 2.3 kV/m is a remnant of the charge left by the emission of the initial particles.

The collision energy also saturates in a similar fashion to other MP parameters (see Fig.20). It is important to notice that RF field strength does not change the saturation level of collision energy. This absence of expected dependence will be discussed later.

### C.  Phase focusing

As pointed out by many authors ([1,11] for example) Without space charge effect the multipacting has a polyphase dynamic at the early stages of development. The polyphase regime is a non-resonant form of electron multipactor and it is characterized by an uniform distribution in time of colliding particles. Fig. 21 shows the emission and collision currents vs time in the simulations without space charge effect. The emission current is deeply modulated because the energy of collision is modulated by the RF field, whereas the colliding particles are distributed almost evenly due to the initial velocity spread as it is supposed to be in polyphase regime. This continuous distribution (though not necessarily strictly uniform) of collision current in the simulations without space charge is also shown in Fig.21.

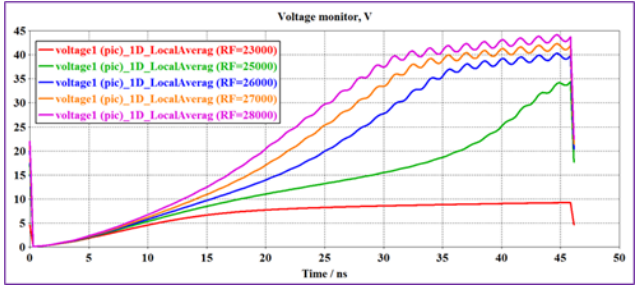

FIG. 19. The voltage monitor readings at different amplitude of RF field.

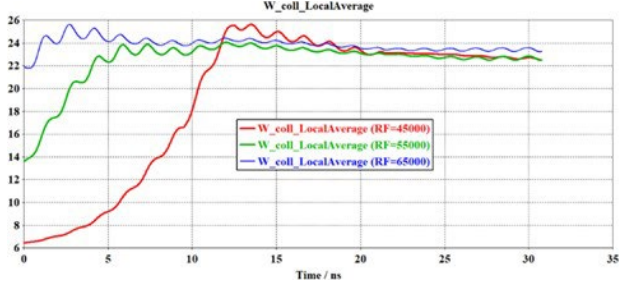

FIG. 20. Saturation of average collision energy at different amplitude of RF field.

In contrast to that the simulations with space charge effect clearly demonstrate phase focusing effect, which is a key property of resonant motion. Fig.22a shows both the colliding and the emission currents coming to saturation and exhibiting deep modulation. It is important to note that at the beginning of the process when the electrostatic field is still weak there is a polyphase stage of multipactor (see the inserts in 22a), though the emission current is already noticeably modulated by the RF field. The transition from polyphase multipacting to resonance is accompanied by a decrease in the amplitude of the average collision energy modulation and saturation of the induced electrostatic field as shown in Fig.22b.

One more important difference between simulations with and without space charge is that without space charge the functions are synchronous, while with space charge they have phase shifts relative to each other (see Fig.23). As a first step in understanding the phase focusing mechanism one can consider that fact.

Fig.24 shows one period of normalized emission current

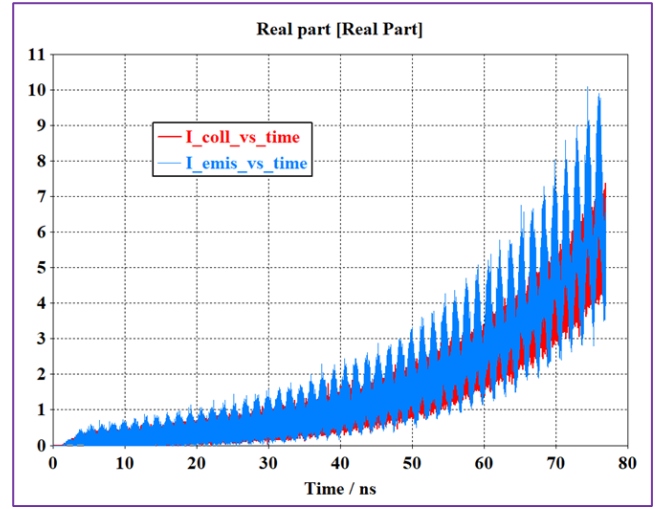

FIG. 21. Collision (in red) and emission (in blue) currents vs time in the simulations without space charge, $E_{dc}$ = 13 kV and $E_{rf}$ = 30 kV.

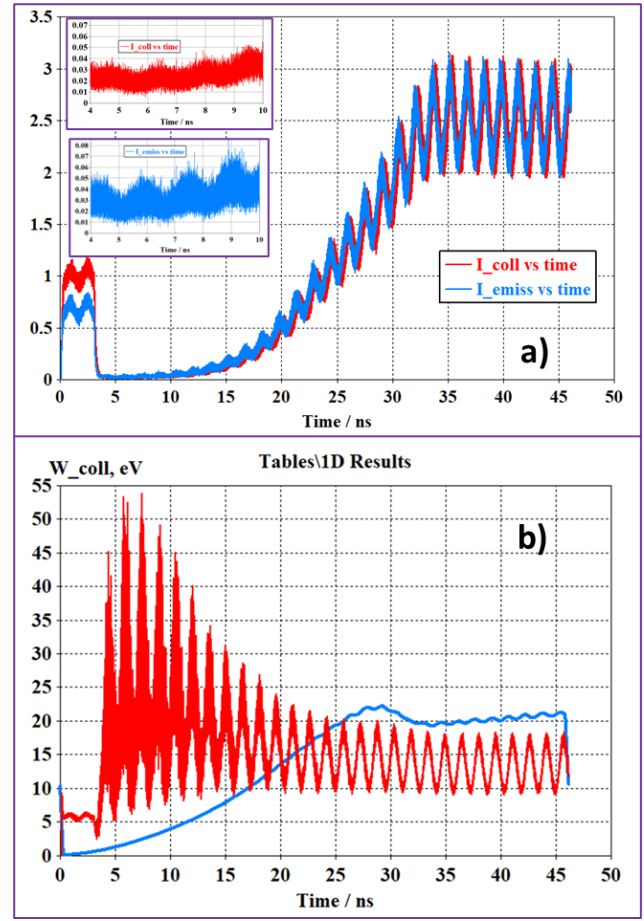

FIG. 22. a) Collision (in red) and emission (in blue) currents vs time in the simulations with space charge effects, $E_{dc}$ is floating and saturates at the level of $E_{dc} \approx$ 12 kV, $E_{rf}$ = 30 kV. The inserts show polyphase regime of both currents in time interval from 4 to 10 ns. Fig.22b shows saturation of the collision energy (in red) with decreasing of the spread and saturation of the floating $E_{dc}$ (in blue, qualitative curve because of uncertainty of the voltage monitor).

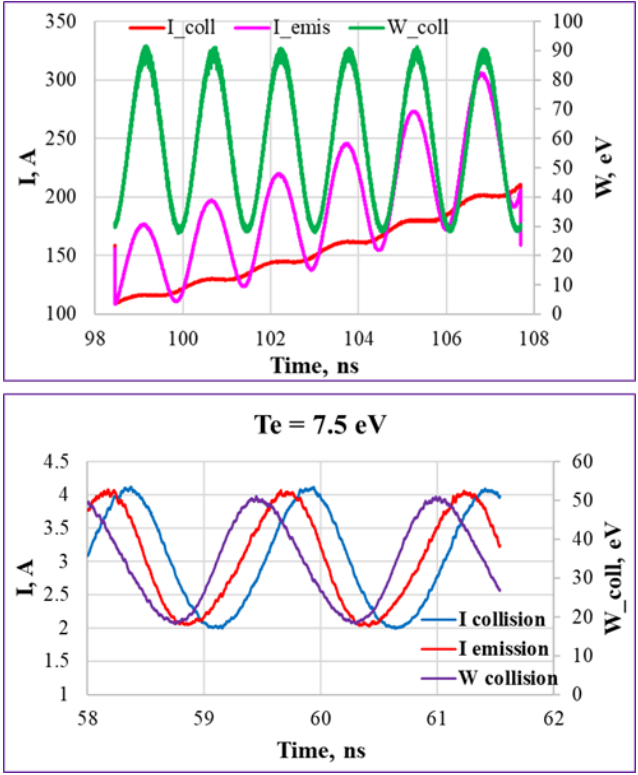

FIG. 23. a) Collision and emission currents and collision energy vs time in simulations without (upper plot) and with space charge effect (bottom plot).

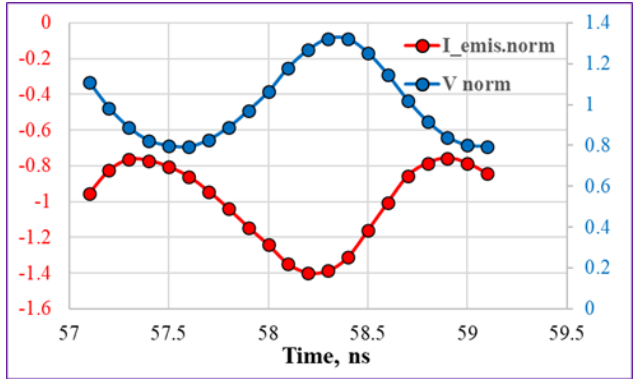

FIG. 24. Normalized emission current through the plane of dielectric surface and normalized readings of the voltage monitor.

through the plane of the dielectric surface and normalized voltage indicated by the voltage monitor (in the chosen coordinate system (Fig.1) the emission current is negative). The voltage monitor is placed on the side of dielectric where MP does not take place, but it reflects the charge of dielectric and the electrostatic field in the vicinity of the dielectric surface. We are interested in this vicinity because this is where the critical modulation of emitted particle velocity happens. The modulated electrostatic field modulates in turn the velocity of all particles and thus their time of flight between emission and collision. Assuming that the electric field of the charged di-

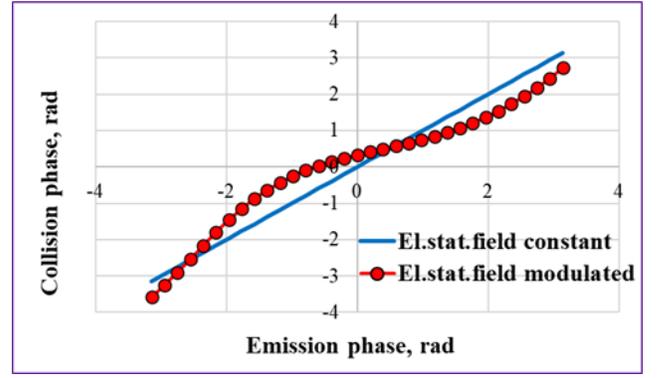

FIG. 25. Phase focusing of the resonant particles. Initially uniform phase spread during emission shrinks after flight due to the initial velocity modulation.

electric is uniform and modulated in time:

$$m\ddot{y} = -eE_{dc}[1 + \alpha \sin(\omega t + \varphi)] \quad (5)$$

where $E_{dc}$ is electric filed amplitude, $\alpha$ – depth of modulation, $\omega$ – frequency of modulation, $\varphi$ – phase of particle emission in relation to the electrostatic field modulation.

The function $V_{norm}(t)$ shown in the plot in Fig.24, is not, of course, a cosine function, and the electric field is not uniform. Nevertheless, let us use (5) to evaluate at least a qualitative effect of the velocity modulation in the given time interval. All particles are resonant with initial velocities corresponding to (3) with a uniform spread in time of emission. Phases of collision as a function of emission phase with and without field modulation are shown in Fig.25, and the phase focusing effect is clearly seen with the simplifications made. Further away from the dielectric the space charge of the particle cloud dominates and that requires, of course, an additional more accurate study of the phase focusing effect.

### D. Synchronous phase

There is one more important feature of the simulations of MP with space charge: while the saturated collision energy does not depend on the amplitude of applied RF field, the phase of collision does depend on the applied RF field strength, which is shown clearly in Fig .26.

Actually the fact that the average collision energy at saturation depends only weekly on the RF field level can be explained by this variation in the initial phase of resonant particle. Fig.27 shows the average initial phase which provides different fixed average energy of collision. These curves are from the simulations and compared to analytical calculations at the same energy for a resonant particle (time of flight is T/2) and variable initial phase using (1).

Fig.28 illustrates the mechanism of the initial phase variation of the resonant particles. The solid lines show energy of collision of the resonant particles vs their initial phase for a given RF field. The collision energies of the individual resonant particles were calculated using (1). The time of flight is always T/2, but depending on the initial phase the particles are accelerated during part of flight and decelerated during the remaining time, excluding initial phases 0° and 180° (the particles are accelerated all the time). Therefore, the energy of collision is the difference between gained and lost energy during the flight. The dashed lines are the crossover energies $W_1$ and $W_2$ of SEY, their levels here are arbitrary and chosen to fit the plot conveniently. Formally MP can start at an RF field level of $E1_{rf}$ once the energy of collision of resonant particles with initial phases of 0° or 180° reaches the first crossover $W_1$. From this point of view, it is clear why the RF field level, at which MP starts, depends on $W_1$. But the number of resonant particles that emit exactly at 0° or 180° is too small to support multiplication, so the MP threshold typically is higher, and MP starts when the range of appropriate initial phases $\theta$ and therefore the number of emitted particles is large enough for the multipacting process to develop, say at a field level $E2_{rf}$ and a corresponding phase interval $\Delta\theta_2$. The multipactor continues with increasing RF field increase as long as $\Delta\theta$ stays sufficiently large. With further increase in RF field $\Delta\theta$ begins shrinking, and presumably MP should stop when $\Delta\theta$ is below some critical value. That level of RF field was not reached in the simulations due to the solver overloading, so this speculation is not verified.

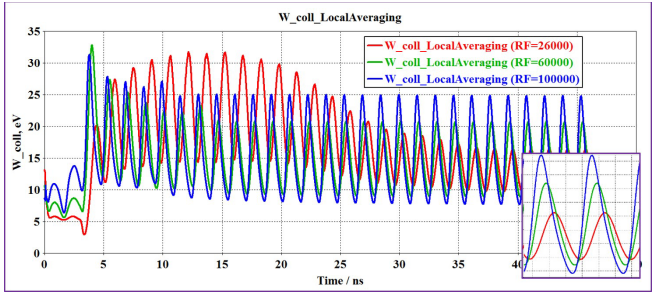

FIG. 26. The collision energies at different levels of RF field are reaching saturation. The insert shows the phase shift of collisions.

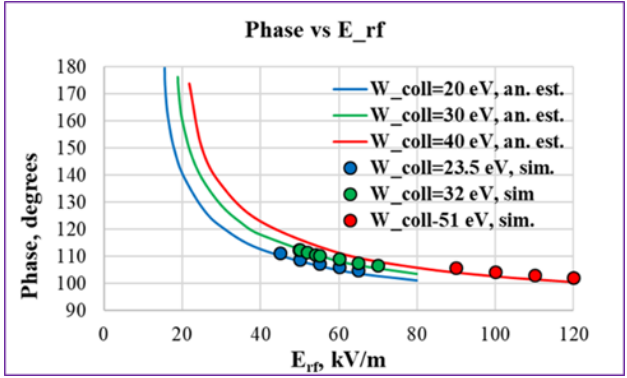

FIG. 27. Average initial phase $\theta$ of the resonant electrons that provides certain $W_{coll}$ at given $E_{rf}$. The phases calculated from the simulations (dots) are compared to the analytical calculations using (1) (lines).

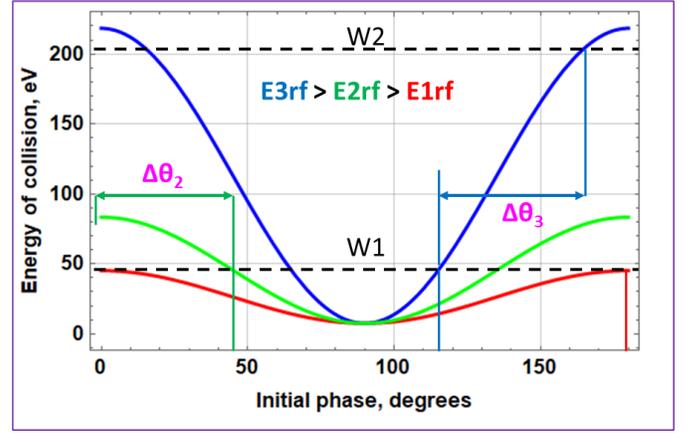

FIG. 28. Figure 28: Analytical evaluation of the resonant particles' energy of collision as a function of initial phase at different amplitudes of RF field. The crossover energies W1 and W2 are arbitrary and serve for qualitative explanation of the initial phase variation. Make green delta theta for E2rf

## V. IMPACT OF THE EMISSION PARAMETERS

Four different sets of SEY functions were used to study the impact of SEY parameters on the multipactor dynamics. The parameters are shown in Tab.1. The initial energy distribution of the secondary particles was the same in all simulations (PDF is shown in Fig.2). Among the emission model parameters, the first crossover $W_1$ of the SEY function plays an especially important role in the MP process. It defines the RF field level at which multipactor begins (threshold) and influences the saturation levels of multipactor parameters. Fig. 29 shows a typical change in collision current and collision energy for different first crossovers.

TABLE I. Sets of secondary emission yield parameters

| $SEY_{max}$ | $W_{max}$, eV | $W_1$, eV | $W_2$, keV |
|---|---|---|---|
| 3 | 200 | 11 | 6.6 |
| 3 | 200 | 16 | 6.5 |
| 1.8 | 150 | 22 | 1.1 |
| 3 | 220 | 31 | 7.0 |

Fig.30 and 31 show the complete set of the collision currents and energies vs RF field amplitude. Apparently, the growth rates of the collision currents are correlated with the second crossover $W_2$.

Both the threshold RF field amplitude and the collision energy at the threshold are linear functions of the first crossover $W_1$ as shown in Fig.32. The theoretical prediction of threshold using (4) is also shown fro comparison. There is a disagreement between the theory and the simulations, and it increases dramatically with $W_1$. The theory assumes a polyphase regime at low $E_{dc}$, and it assumes also that it remains polyphase. But the voltage (i.e. $E_{dc}$) sharply jumps to a much higher level at threshold (Fig.33). This means that the electrons with higher initial energy $W_0$ and therefore more nu-

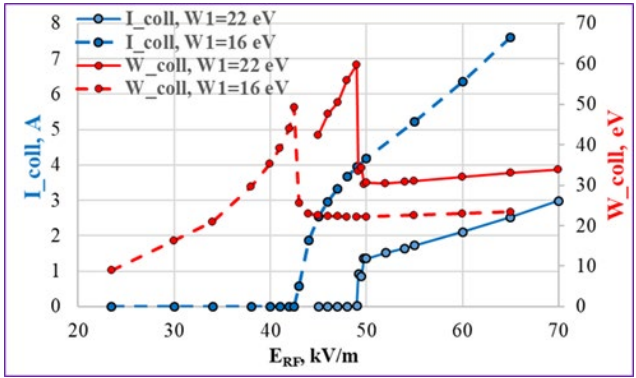

FIG. 29. Average collision current $I_{coll}$ and collision energy $W_{coll}$ of multipactor as functions of the RF field amplitude.

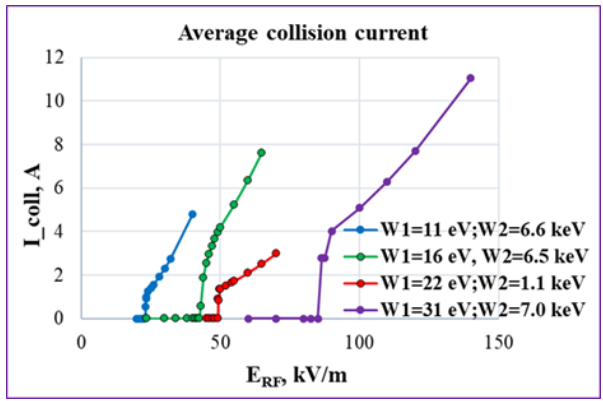

FIG. 30. The collision currents vs RF field amplitude for different crossovers $W_1$ and $W_2$.

merous become resonant, so the overall MP process becomes dominantly resonant

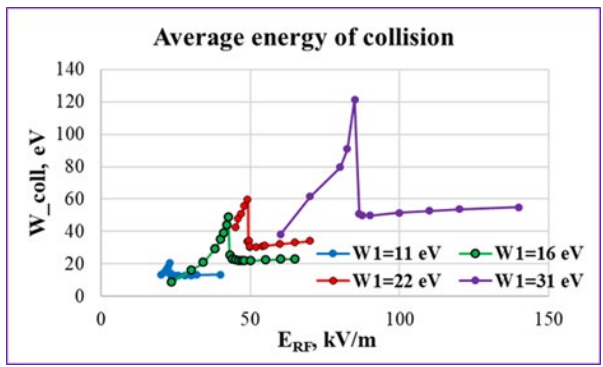

FIG. 31. The collision energies vs RF field amplitude for different first crossovers W1.

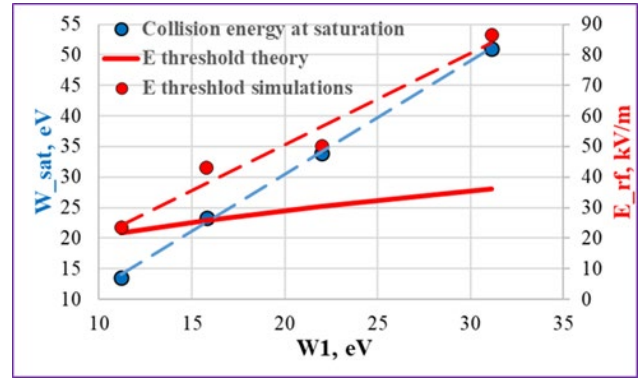

FIG. 32. RF field threshold and average collision energy at saturation vs first crossover of SEY. The threshold according to theory [2], which assumes completely polyphase regime, is given for comparison.

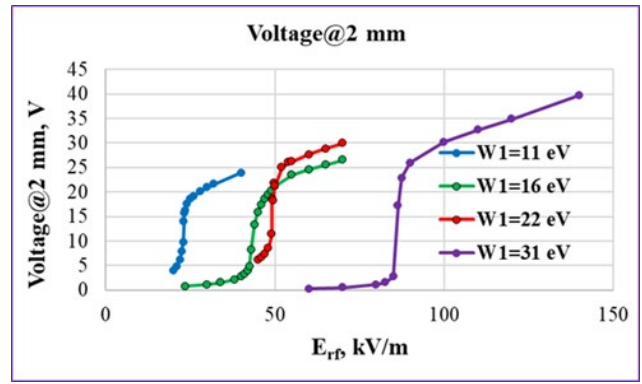

FIG. 33. The voltage monitor readings vs RF field amplitude and different first crossovers of SEY functions.

## VI. SUMMARY

This work is to analyse the results of multi-particle PIC simulations of one-side multipactor on dielectric, performed with space charge effects included. The observations and analysis of the multipacting process features have led to the conclusion that the one-side multipactor on dielectrics is essentially a resonant process. The following list summarizes the main results of the study.

- The initial energy of secondary electrons is a secondary emission property of utmost importance, because essentially it is a base for the resonant dynamic of one-side multipacting on a dielectric. With any charge on the dielectric (i.e. returning electrostatic field level) there is always a synchronous secondary electron with time-of-flight T/2 due to the continuous distribution of initial velocities.

- The non-uniform distribution of initial energies/velocities of secondary particles (Maxwellian in our case) is of fundamental importance too. The charging process adjusts the charge on the dielectric and therefore the strength of returning electrostatic field in such a way that secondary electrons with initial energy equal or close enough to the maximum of PDF are resonant. Therefore the charge on dielectric is floating and limited for a given set of parameters.

- The resonant discharge does not start immediately after the threshold of $E_{rf}$ is exceeded. There are two overlapping stages of MP development: 1) the dominantly polyphase regime in the beginning of MP and 2) the dominantly resonant regime at saturation.

- The collision current is modulated in the simulation with space charge effects active, which allows the assumption that some phase focusing occurs. The charge on the dielectric and the returning electrostatic field oscillate at saturation. The preliminary simplified view of particle dynamics shows that the modulated returning field provides phase focusing.

- There is no strong dependence of average collision energy on RF field level, it is almost constant for a given first crossover $W_1$. It was found that the MP dynamics keeps the average energy of collision $W_{coll}$ constant via variable synchronous phase while the RF field is increasing.

- The interval of favorable initial phases ( the phases at which the resonant particles gain energy of collision which satisfies the condition $W_1 < W_{coll} < W_2$) starts decreasing above some RF field level. This presumably can define the upper threshold for the RF field, but this level was not reached and confirmed in the simulations due to technical issues.

Some of the effects of resonance multipacting on dielectrics revealed by PIC simulations, such as the phase dispersing effect of the space charge or the impact of RF field frequency, still require more detailed considerations. For completeness, results on the latter problem are presented in the Appendix.

**Appendix A: Simulation results for different frequencies**

Simulations of one side multipactor on dielectric at different frequencies.

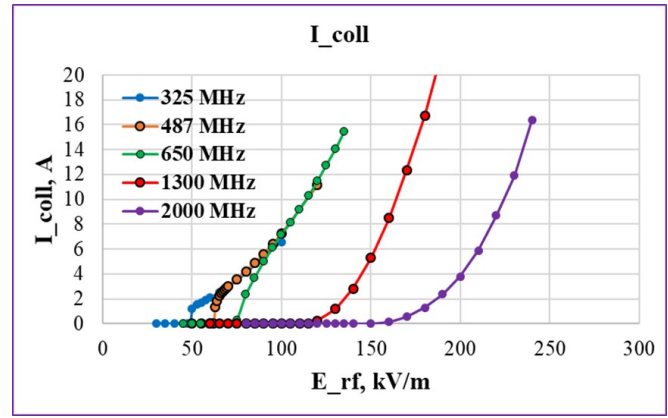

FIG. 34. Shift of the RF electric field thresholds for different frequencies.

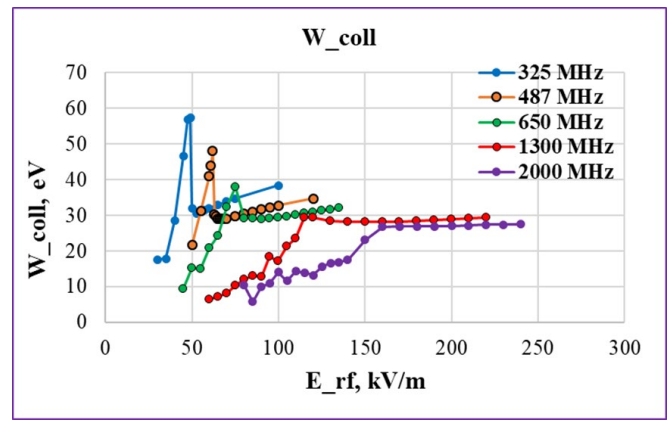

FIG. 35. Collision energies vs RF field strength for different frequencies.

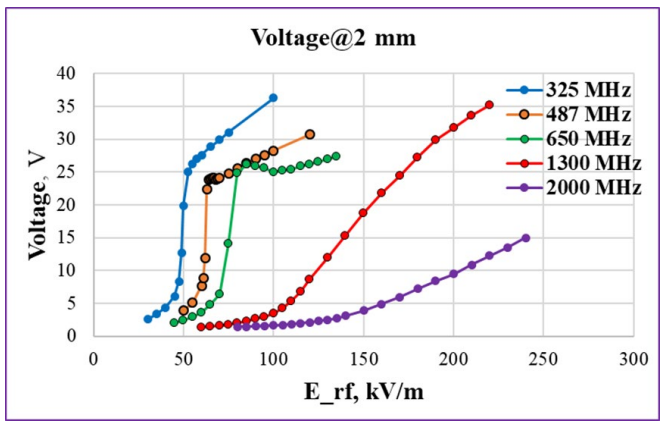

FIG. 36. Voltage monitor reading vs RF field strength for different frequencies.